\documentclass[12pt]{article}

\usepackage{newtxtext,newtxmath}

\usepackage{graphicx}

\usepackage[letterpaper,margin=1in]{geometry}

\usepackage{amsmath}

\renewenvironment{abstract}
	{\quotation}
	{\endquotation}

\date{}

\makeatletter
\renewcommand{\fnum@figure}{\textbf{Figure \thefigure}}
\renewcommand{\fnum@table}{\textbf{Table \thetable}}
\makeatother

\usepackage{scicite}

\usepackage{url}

\def\scititle{
	Observing Laughlin's pump using quantized edge states in graphene
}
\title{\bfseries \boldmath \scititle}

\author{
	Bjarke~S.~Jessen$^{1,2\ast\dagger}$,
	Maëlle~Kapfer$^{2\dagger}$,
	Yuhao~Zhao$^{3\dagger}$,\and
	Kenji~Watanabe$^{4}$,
	Takashi~Taniguchi$^{5}$,
	Cory~R.~Dean$^{2}$,
	Oded~Zilberberg$^{6\ast}$ \\
	\small$^{1}$Department of Physics, Technical University of Denmark, Kgs. Lyngby 2800, Denmark\and
	\small$^{2}$Department of Physics, Columbia University, New York 10027, USA\and
	\small$^{3}$Institute for Theoretical Physics, ETH Zurich, Zurich 8093, Switzerland\and
	\small$^{4}$Research Center for Electronic and Optical Materials, National Institute for Materials Science,\\[-0.8em]\small 1-1 Namiki, Tsukuba 305-0044, Japan\and
	\small$^{5}$Research Center for Materials Nanoarchitectonics, National Institute for Materials Science,\\[-0.8em]\small 1-1 Namiki, Tsukuba 305-0044, Japan\and
	\small$^{6}$Department of Physics, University of Konstanz, Konstanz 78464, Germany\and
	 \small$^\ast$Corresponding author. Email: bsoje@dtu.dk\and
	\small$^\ast$Corresponding author. Email: oded.zilberberg@uni-konstanz.de\and
	\small$^\dagger$These authors contributed equally to this work.
}

\begin{document} 

\maketitle

\begin{abstract} \bfseries \boldmath
Laughlin’s thought experiment of quantized charge pumping is central to understanding the integer quantum Hall effect (IQHE) and the topological origin of its conductance quantization. Its direct experimental observation, however, has been hindered by the difficulty of realizing clean electronic edges. We address this by fabricating ultra-small, lithographically defined contacts on graphene. This creates a Corbino-equivalent system, with well-confined inner edge states. Crucially, the small contact size induces strong energy quantization of the edge states.
This quantization allows us to directly resolve the spectral flow associated with Laughlin's pump. By tracing the finite-size resonances of the inner edge, we observe clear oscillations in conductance as a function of magnetic field and carrier density. The oscillation period scales with contact size, consistent with quantized charge transfer.
Thus, our results provide a direct observation of the spectral flow underlying Laughlin's pump.
The simplicity of the graphene platform makes this approach scalable and robust for exploring fundamental topological effects.

\end{abstract}

\noindent
The discovery of the integer quantum Hall effect (IQHE) revealed that a two‑dimensional electron system under a perpendicular magnetic field has its Hall conductance quantized in exact integer multiples of $e^{2}/h$, where $h$ is Planck's constant and $e$ is the electron charge~\cite{klitzing:1980}. This quantization was later shown to originate from the topological nature of electron wavefunctions in a magnetic field~\cite{PhysRevLett.49.405, PhysRevLett.51.51, Q_Niu_1984}. While the quantized Hall conductance has been confirmed both experimentally and theoretically, direct observation of the underlying physical mechanism proposed by Laughlin has remained elusive~\cite{laughlin:1981}.
Laughlin's thought experiment considers a 2D electron system rolled into a cylindrical geometry under an out-of-plane magnetic field. An adiabatic insertion of a single magnetic flux quantum $\phi_0 = h/e$ through the cylinder's axis, then, pumps an integer number of electrons from one edge of the system to the other. In the topologically equivalent Corbino disk geometry, this charge transfer appears as a quantized radial current induced by the flux insertion~\cite{PhysRevB.25.2185}, see Fig.~\ref{fig:1}A. 
The idea is theoretically elegant, and its mechanism has been explored in bosonic emulators such as photonic and cold‑atom analogues~\cite{PhysRevLett.109.106402, PhysRevB.91.064201,hafezi_measuring_2014, lohse2016thouless, ke2016topological}. However, direct observation of quantized electron charge transfer in mesoscopic devices remains elusive. Disorder, local imperfections, and the difficulty of fabricating pristine edges have masked the effect in several device architectures~\cite{weis2011metrology,kou2012coulomb,camino2005aharonov,mcclure2012fabry}.

In this work, we realize Laughlin’s thought experiment using a graphene-based electronic system. The device incorporates a small, lithographically defined contact, fabricated with a recently developed technique for high-quality interior contacts~\cite{zhao2024emergent}. This simple architecture creates well-defined electrostatically defined edges, and realizes a Corbino-like geometry with spatially separated interior contacts. We measure charge transport through the bulk in a perpendicular magnetic field. By reducing the size of the interior contact, we confine the edge states, leading to Bohr-Sommerfeld quantization. This finite-size quantization allows us to resolve the energetically isolated edge states and directly trace the spectral flow of the charge pumping argument.
Specifically, when the magnetic field or Fermi level is varied, we observe pronounced oscillations in the conductance. The oscillation periods, $\Delta B=\phi_0/(\pi R^2)$ and $\Delta \varepsilon = \tilde{v}_F \hbar/R$, with $\tilde{v}_F$ being the edge-state velocity, and $\hbar$ the reduced Planck's constant, matches the expected spacing between the quantized edge states encircling the contact of radius $R$. Our results thus provide a direct realization of Laughlin’s thought experiment and reveal the underlying spectral flow of edge states responsible for quantized pumping.
The simplicity of our scheme enables direct extraction of key edge-state properties such as velocity and energy spacing.
Graphene's high carrier mobility and linear dispersion near charge neutrality make it a versatile platform for probing and manipulating topological edge states. This approach offers a new route to explore quantum Hall physics and topological phases in electronic systems.

\subsection*{Size-quantized edge states}
\begin{figure}[th]
	\centering
	\includegraphics[width=1\linewidth]{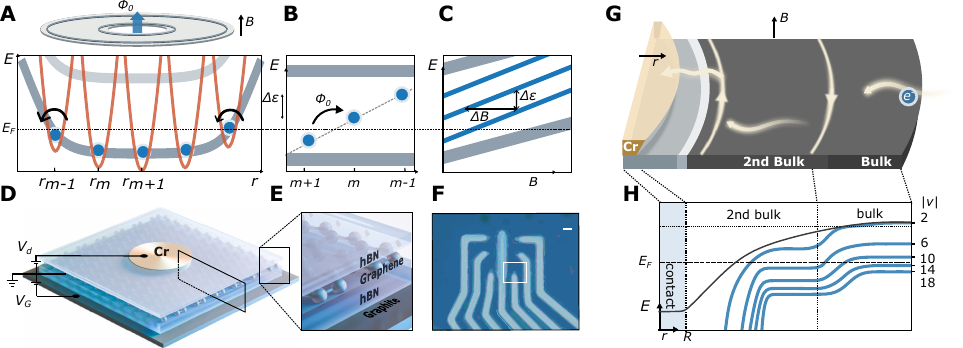}%
\caption{\textbf{Laughlin's pump using a small top contact on graphene.} (\textbf{A}) Upper panel: Illustration of Laughlin's pump in a Corbino disk geometry. Threading a magnetic flux through the hole in the disk,  radially pumps electrons  outwards~\cite{laughlin:1981}. Lower panel: Electrons (blue dots) fill their respective guiding centers (orange lines). The Landau levels (gray lines) bend in the radial direction at the system edges. Threading of a flux quantum, $\phi_0$, shifts the guiding centers by one $r_{m+1}\rightarrow r_m$, and pushes charges across the Fermi energy $E_F$ at the edges.  (\textbf{B}) Dispersion of the edge states (thin gray line) as a function of the angular wavenumber $m$. For a small edge circumference, the edge dispersion is sparsely sampled (blue dots).   (\textbf{C}) Spectral flow of the quantized edge states as a function of the magnetic field. (\textbf{D-E}) Illustration of the top contact and device. The detailed structure in (\textbf{D}) shows the architecture of the top contact, as marked in the false-colored SEM photo shown in (\textbf{F}). 
The length bar in (\textbf{F}) corresponds to $1\,\mathrm{\mu m}$. (\textbf{G}) Illustration of the electron transport in real space. In the magnetic field, the top contact induces a secondary graphene bulk~\cite{zhao2024emergent}. Lateral transport is through the edge states.  (\textbf{H}) The top gate defines a smooth radial doping profile (black line). In a magnetic field, it reconstructs into a wedding cake potential~\cite{gutierrez_interaction-driven_2018,zhao2024emergent}, supporting two sets of LLs (blue lines).}
	\label{fig:1}
\end{figure}

We consider a two-dimensional electron gas in a Corbino disk geometry under a perpendicular magnetic field $B$, see Fig.~\ref{fig:1}A. Exploiting the rotational symmetry of the system, we express the electron wavefunction as $\psi(r,\varphi)=\psi(r)e^{im\varphi}$, where $m$ is a quantized angular wavenumber. For each $m$ the radial motion of electrons is governed by the Hamiltonian
\begin{equation}
H_m=\frac{\hbar^2}{2m_e}\left[r^{-1}\partial_r (r\partial_r) + \left(\frac{m}{r}-\frac{r}{2{l_B}^2}\right)^2\right],
\end{equation}
where $\hbar$ is the reduced Planck constant, $m_e$ the electron effective mass, and ${l_B}=\sqrt{\hbar/|eB|}$ is the magnetic length. For each $m$, the electrons experience the same confinement in the radial direction, but are centered around a different guiding center $r_m=\sqrt{2m}l_B$. In the bulk, the confinement engenders Landau levels (LLs), i.e. a discrete set of energy levels that are highly degenerate due to the abundance of $m$. Electrons with guiding centers at the edge of the device experience additional confinement that (de)increases their energy, leading to the well-known edge states of the QHE~\cite{Laughlin:1983}. The chemical potential (Fermi level) of the system determines which LLs and guiding centers are filled, see Figs.~\ref{fig:1}A and 
B.

Adiabatically threading a magnetic flux $\phi$ through the central hole of the Corbino geometry radially shifts the guiding centers $r_m$. An insertion of precisely one flux quantum $\phi_0=h/e$ moves electrons from the guiding center at $r_m$ to the next location at $r_{m-1}$. As flux is threaded, each guiding centre shifts radially. Levels that cross the chemical potential at the inner edge release an electron, while those at the outer edge absorb one. This spectral flow \cite{laughlin_1981}, together with Faraday’s law, makes up Laughlin’s charge‑pump argument and yields the integer‑quantised Hall conductance \cite{Laughlin:1983, PhysRevB.25.2185}.

When the inner edge has a small circumference $2\pi R$, the possible values of  $m$  are restricted by Bohr-Sommerfeld discretization, and a continuous spectral flow is thus not supported. Assuming a constant group velocity $\tilde{v}_F$ along the edge, the discrete values of $m$ imply an energy spacing $\Delta \varepsilon = \tilde{v}_F \hbar/R$ at the edge, see Fig.~\ref{fig:1}B. The smaller $R$, the larger the spacing $\Delta \varepsilon$. Each discrete angular mode $m$ represents a standing-wave state analogous to modes in a one-dimensional ring~\cite{buttiker1983josephson}. When such a state is energetically close to the Fermi level, it acts as a resonant scattering channel, enhancing electrons' transport to the nearby contact~\cite{ihn_semiconductor_2009}. Hence, by tuning the chemical potential, we can cross  discernible isolated transport resonances associated with the discrete inner-edge states~\cite{ihn_semiconductor_2009} in experiments. 
On the other hand, these resonances are tunable through the magnetic field due to the aforementioned spectral flow~\cite{hafezi_measuring_2014}, see Fig.~\ref{fig:1}C. Here, the invariance of the system to the insertion of a flux quantum implies that the period between resonances as a function of $B$ is $\Delta B=\phi_0/\pi R^2$. Combined, the discretized spectral flow reads $\varepsilon\approx (\Delta \varepsilon/\Delta B)B + m\Delta \varepsilon$, which leads to the resonance condition
\begin{equation}\label{eq: edge states}
\frac{1}{\Delta \varepsilon}\varepsilon - \frac{1}{\Delta B} B=m\,.
\end{equation}
This condition predicts a series of resonances to appear in an experiment as oscillations in conductance across the device. Yet, challenges in fabricating clean and small $2\pi R$ inner edges in a Corbino disk geometry hindered the measurement of this effect, and the direct observation of Laughlin's argument in a condensed matter setting~\cite{zeng_high-quality_2019, yan2010charge, polshyn_quantitative_2018}.

\subsection*{Experimental Realization of a clean inner edge}
To achieve a clean and well-defined inner edge, we employ a recently developed fabrication method that creates sub-micrometer, high-quality interior metal contacts directly on top of encapsulated graphene~\cite{zhao2024emergent}. This approach utilizes a gentle, metal-assisted defluorination process, resulting in a pristine graphene-metal interface with minimal disorder or edge roughness, see Figs.~\ref{fig:1}D-F. 
To fabricate the interior contacts, we encapsulate monolayer graphene between two hexagonal boron nitride (hBN) crystals and place the stack on a graphite back-gate. Using fluorine-based dry etching, a small window is then etched into the top hBN layer to locally expose the graphene surface. Using the same mask, a metal trilayer (Cr/Pd/Au) is subsequently deposited into the etched region, defluorinating and forming a high-quality nanoscale contact directly on the graphene \cite{zhao2024emergent, son2018atomically, kretz2018atomistic, jessen2019lithographic}. This approach yields well-defined electrostatic edges with minimal disorder, essential for resolving the quantized edge states in our measurements.
Crucially, the presence of these metal contacts electrostatically dopes the underlying graphene, effectively defining a pristine inner edge for the Corbino geometry. In the presence of a perpendicular magnetic field and strong electron-electron interactions, this electrostatic doping profile evolves into a characteristic ``wedding-cake'' potential~\cite{gutierrez_interaction-driven_2018,zhao2024emergent}, illustrated in Figs.~\ref{fig:1}G and H. This structured potential supports two distinct sets of Landau levels (LLs), associated with spatially separated regions of different doping—referred to as the ``main bulk'' and the ``secondary bulk.'' As the LLs supported by the latter are located closest to the contact, they experience the smallest circumference of the inner edge and hence yield discernible resonance signals in the transport measurements below.

\subsection*{Observation of Laughlin's charge pumping}
In Fig.~\ref{fig:2}A, we show the two-terminal conductance, $G$, measured for a device with a $100\,\mathrm{nm}$ radius contact as a function of a perpendicular magnetic field $B$ and charge carrier $n_s$. At high magnetic fields ($B>2\,\mathrm{T}$), the conductance reveals two distinct, intersecting sets of Landau fans. To clearly identify conditions under which each bulk region is gapped, we plot the data against the filling factor of the main bulk, $\nu=n_s\phi_0/B$, see Fig.~\ref{fig:2}B. 
Notably, when the secondary bulk is insulating while the main bulk remains conducting, we observe pronounced oscillations in $G$, see inset of Fig.~\ref{fig:2}B.
To quantitatively verify the periodicity of these oscillations, we extract line cuts of conductance at fixed filling factor $\nu$ and fit them to Eq.~\eqref{eq: edge states}, where we associate filling to energy via $\nu=\frac{\varepsilon^2}{v_{F}^{2}}\frac{2}{eB\hbar}+2$. The resulting fit (Fig.~\ref{fig:2}B, inset) yields excellent agreement with theory for an effective inner edge radius of $R=105\,\mathrm{nm}$ and edge-state velocity ratio $v_F/\tilde{v}_F=2$, relative to graphene’s intrinsic Fermi velocity ($v_F=10^6\,\mathrm{m/s}$). The slight discrepancy from the nominal $100\,\mathrm{nm}$ lithographic radius is attributed to the contact geometry and associated electrostatic environment, which slightly increases the effective edge circumference (see Extended Data).

\subsection*{Multiple edge states}
Thus far, we have neglected the effect of filling factor on the observed conductance oscillations. As such, at first glance, the excellent agreement between our simplified theory and the experiment might appear surprising: according to the bulk-boundary correspondence, the number of inner edge states should match the filling factor of the secondary bulk, $\nu_2$. Due to Pauli’s exclusion principle, each of these $\nu_2$ edge states occupies slightly different radii, resulting in small energy shifts $\delta \varepsilon$ between them. These shifts further lead to different resonance conditions and distinct quantized energy spacings $\Delta\varepsilon_i$ at the inner edge, where $i=1,\ldots,\nu_2$, see Fig.~\ref{fig:2}C. Correspondingly, one would expect $\nu_2$ distinct spectral flows as a function of the magnetic field $B$~\cite{PhysRevB.46.4026}, see Fig.~\ref{fig:2}D. This suggests that Eq.~\eqref{eq: edge states} should be insufficient to describe the observed conductance oscillations.
However, such a refined description turns out to be unnecessary for our experimental conditions, owing to both the inherently small energetic offset $\delta \varepsilon$ and the finite measurement bandwidth $\Gamma$, which arises primarily from thermal broadening and moderate disorder. 
Specifically, $\Gamma$ is narrow enough to clearly resolve the average inner-edge energy spacing $\Delta \bar{\varepsilon}=(1/\nu_2)\sum_{i=1}^{\nu_2}\Delta \varepsilon_i$, yet broad enough to smear out the finer variations $(\delta \varepsilon)^2/\Delta\bar{\varepsilon}$ due to the slightly different radii of individual inner edges. 
To validate this interpretation, we numerically calculate the longitudinal conductance across a Corbino-disk-like system, see Figs.~\ref{fig:2}E, F. The oscillation pattern identified in Fig.~\ref{fig:2} confirms that the appearance of edge states modulates the transport signal. Moreover, already by introducing moderate-level disorder, the fine oscillation pattern expected from variable circumference inner edges is smeared out to coincide with our effective averaged ``single-edge'' model. 

\subsection*{Scaling with Corbino hole radius and robustness to temperature increase}
We demonstrated Laughlin's pumping argument using a small metal contact on graphene. We relied on the small circumference of the contact to resolve the inner edge's quantized energy spacing $\Delta\varepsilon_i\propto 1/R$. Similarly, from Eq.~\eqref{eq: edge states}, the period of the oscillation in the magnetic field is $\Delta B \propto 1/R^2$. To verify this dependence on $R$, we perform the same measurements on devices with varying contact radii, $R=100,150,200\,\mathrm{nm}$, see Figs.~\ref{fig:3}A and B. We find clear oscillation patterns in all three devices, with increasing period for decreasing contact radii as expected. Furthermore, we extract the mean period $\Delta B$ for each device and show it as a function of the contact radius in Fig.~\ref{fig:3}B. The observed scaling agrees well with the expected $\propto 1/R^2$ scaling. This provides further quantitative evidence for the size quantization of the inner edge states and  verifies our approach for probing Laughlin's pumping mechanism. Moreover, we find that the measured oscillations survive at elevated temperatures up to $10\,\mathrm{K}$, see  Fig.~\ref{fig:3}C. This implies that the thermal broadening $\Gamma_T$ suffices to resolve the energy spacing $\Delta \varepsilon$, whereas the period $\Delta B$ is ensured by the topological nature of Laughlin's pumping. Importantly, it also clarifies why the effect remained elusive until now: previous experiments could not simultaneously achieve the necessary combination of low measurement bandwidth (clean, low-temperature conditions) and sufficiently large energy quantization (small contact circumference) to satisfy the condition $\Delta \varepsilon > \Gamma$.

\subsection*{Extracting edge-state dynamics from conductance oscillations}
\begin{figure}[t!]
	\centering
	\includegraphics[width=1\linewidth]{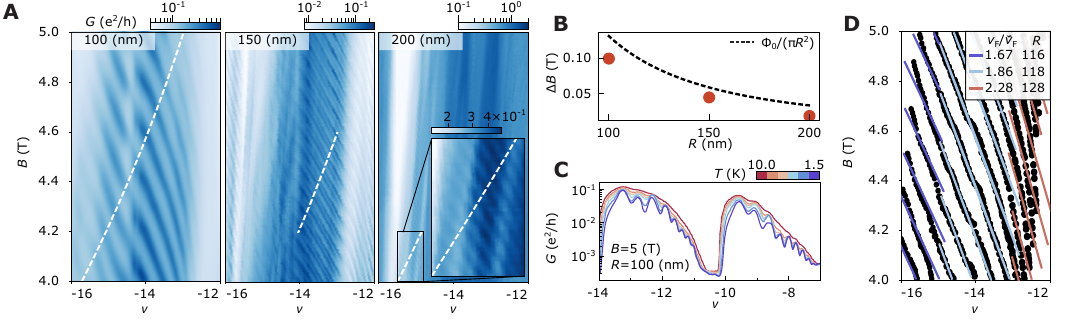}
\caption{\textbf{Validation and parameter extraction: scaling with contact size, temperature robustness, and finding the edge group velocity.}
 (\textbf{A}) Left to right: Two-terminal conductance measured as a function of the filling $
\nu$ and the magnetic field $B$ with  contact radius $100\,\mathrm{nm}$, $150\,\mathrm{nm}$ and $200\,\mathrm{nm}$. (\textbf{B}) The period $\Delta B$ of the oscillations extracted along dashed white lines in (\textbf{A}) (red dots), compared with  Laughlin's argument $\Delta B=\phi_0/(\pi R^2)$ (dashed line), cf.~Supplementary materials. (\textbf{C}) Two-terminal conductance measured at $B=5\,\mathrm{T}$ as a function of filling $\nu$ for several temperatures. (\textbf{D}) By fitting the oscillation pattern in (\textbf{A}) (black dots) with the analytical expression~\eqref{eq: edge states}, we can find the group velocity $\tilde{v}_F$ along the inner edge and its radius $R$  around the $100\,\mathrm{nm}$ contact. The fitting is conducted separately for different ranges of $\nu$, denoted by the red, blue, and purple lines (see also in Supplementary materials).}

\label{fig:3}
\end{figure}

Our device and measurements offer a unique glimpse into the physical properties of the quantum Hall edge states, such as their effective radius and velocity. Specifically, as the conductance oscillations are directly linked to the energy spacing \(\Delta \varepsilon = \tilde{v}_F \hbar / R\), it allowed us to determine the effective edge circumference (\(2\pi R\)) in Fig.~\ref{fig:3}C. Crucially, combining the oscillation data in both filling and magnetic field enables us to extract the velocity \(\tilde{v}_F\) as well as the effective radius $R$ of the inner edge state using the resonance condition Eq.~\eqref{eq: edge states}, see Fig.~\ref{fig:3}D. We find the effective edge radius to be consistently larger than the lithographically defined radius ($R>100\,\mathrm{nm}$), attributed to the electrostatic environment formed around the contact. The extracted edge velocity varies notably, from slower values ($\tilde{v}_F\approx v_F/2.28$) close to Landau level energies to faster values ($\tilde{v}_F\approx v_F/1.67$) deeper in the energy gaps, yet always remains below graphene’s intrinsic Fermi velocity, $v_F$. Such rare access to measuring the edge velocity enhances our understanding of edge-state physics and opens avenues for deeper exploration of quantum Hall systems~\cite{mcclure2009edge, sahasrabudhe2018optimization}.

\subsection*{Conclusion}
Our results confirm the quantized nature of Laughlin’s pump and its robustness against thermal and disorder-induced smearing, establishing a direct link between theory and experiment.
By exploiting the energy quantization of confined edge states, we access key properties of the IQHE, including the effective radius and velocity of edge modes. This method creates new opportunities to study edge dynamics, confinement effects, and the role of disorder and interactions in quantum Hall systems.
Beyond fundamental insights, controlled charge pumping offers a route to precision metrology~\cite{klitzing:2017,RevModPhys.85.1421}, quantum information processing~\cite{RevModPhys.80.1083}, and scalable nanoscale device engineering. 
Looking forward, small metal-on-graphene contacts provide a new platform for electronic interferometry~\cite{wei_mach-zehnder_2017,Biswas_AB_Edge} and for probing interference phenomena in the (anomalous) fractional quantum Hall regime~\cite{henny1999fermionic,hu_high-resolution_2025,lu_fractional_2024}. These advances set the stage for future quantum technologies based on controlled topological transport.



\begin{figure}[t!] 
	\centering
	\includegraphics[width=1\linewidth]{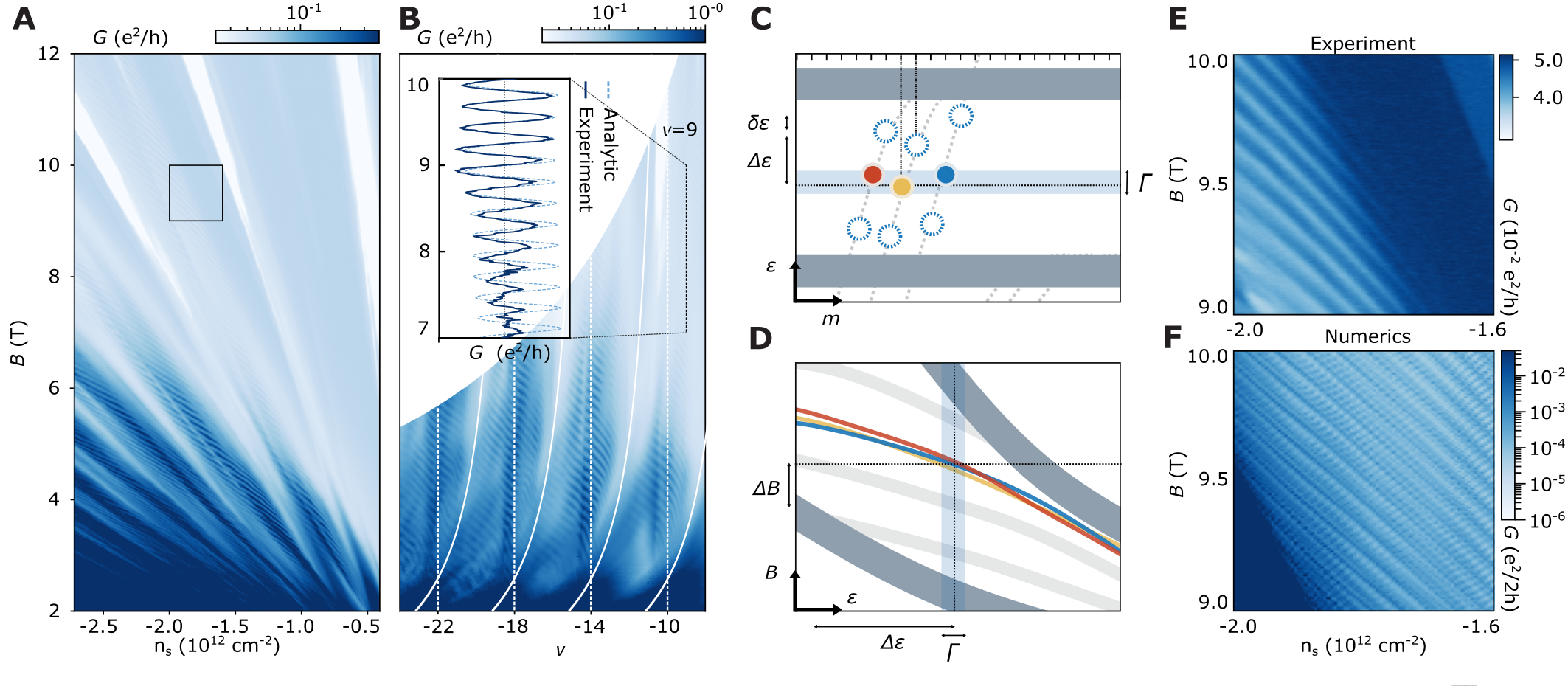}
\caption{\textbf{Transport measurement of Laughlin's pump.} 
(\textbf{A}) Two-terminal conductance measured with a $
100$\,nm-radius contact as a function of a perpendicular magnetic field, $B$, and carrier density, $n_s$. Oscillations appear within the Landau fan of the main bulk when the secondary bulk is gapped. 
(\textbf{B}) Same as (\textbf{A}), plotted as a function of the main bulk's filling $\nu$ and magnetic field. Thus, the LLs of the main bulk are approximately constant at $\nu=4(n+\frac{1}{2})$ with $n\in \mathbb{Z}$ (dashed white lines), and the LLs of the second bulk are curved (solid white lines). Inset: linecut (dark blue) at filling factor $\nu$ = 9 (dotted line), compared with a series of Lorentzians (light blue) fulfilling the resonance condition~\ref{eq: edge states}. Smooth experimental data background is removed for clarity.   
(\textbf{C}) and (\textbf{D}) Same as Figs.~\ref{fig:1}B,C for the multiple edges states that flow between the secondary bulk LLs. Small variations in the quantization conditions of the different edges can lead to variation in the resonances  $\delta\epsilon$.
(\textbf{E}) A zoom-in on the black square in (\textbf{A}). (\textbf{F}) Same as (\textbf{E}), as calculated numerically (see Supplementary materials). The counter-spectral flow is due to finite-size numerical artifacts (the outer edge states are also resolved due to the small circumference). }
	\label{fig:2}
\end{figure}

\clearpage 

%

\bibliographystyle{sciencemag}

%
%
%
%
%
%

\section*{Acknowledgments}
\paragraph*{Funding:}
B.S.J. gratefully acknowledges support from the Villum Foundation.
O.Z. and Y.Z. acknowledge support by the Deutsche Forschungsgemeinschaft (DFG) via project numbers 449653034; 425217212; 521530974; 545605411; and through SFB1432, as well as Eidgenössische Technische Hochschule Zürich research Grant No. ETH-28 23-1.
B.S.J. and M.K. acknowledge support from the Center for Programmable Quantum Materials (Pro-QM), an Energy Frontier Research Center funded by the US Department of Energy (DOE), Office of Science, Basic Energy Sciences (BES), under award DE-SC0019443. 
K.W. and T.T. acknowledge support from the JSPS KAKENHI (Grant Numbers 21H05233 and 23H02052) , the CREST (JPMJCR24A5), JST and World Premier International Research Center Initiative (WPI), MEXT, Japan.
\paragraph*{Author contributions:}
B.S.J. and M.K. conceived of the project.
B.S.J. and M.K. prepared the samples and performed the transport measurements.
Y.Z. and O.Z. developed the analytical model. Y.Z. performed the transport simulations. B.S.J., M.K., and Y.Z. analyzed the data.
K.W. and T.T. grew and provided the \textit{h}BN crystals.
C.R.D. supervised the transport measurements.
B.S.J., Y.Z., M.K., and O.Z. wrote the paper with input from all authors. 
\paragraph*{Competing interests:}
The authors declare no competing interests. 
\subsection*{Supplementary materials}
Materials and Methods\\
Fig. S1\\


\newpage


\renewcommand{\thefigure}{S\arabic{figure}}
\renewcommand{\thetable}{S\arabic{table}}
\renewcommand{\theequation}{S\arabic{equation}}
\renewcommand{\thepage}{S\arabic{page}}
\setcounter{figure}{0}
\setcounter{table}{0}
\setcounter{equation}{0}
\setcounter{page}{1} 


\begin{center}
\section*{Supplementary Materials for\\ \scititle}

	Bjarke~S.~Jessen$^{1,2\ast\dagger}$,
	Maëlle~Kapfer$^{2\dagger}$,
	Yuhao~Zhao$^{3\dagger}$,\\
	Kenji~Watanabe$^{4}$,
	Takashi~Taniguchi$^{5}$,
	Cory~R.~Dean$^{2}$,
	Oded~Zilberberg$^{6\ast}$\\[1.8em]
	\small$^{1}$Department of Physics, Technical University of Denmark, Kgs. Lyngby 2800, Denmark\\
	\small$^{2}$Department of Physics, Columbia University, New York 10027, USA\\
	\small$^{3}$Institute for Theoretical Physics, ETH Zurich, Zurich 8093, Switzerland\\
	\small$^{4}$Research Center for Electronic and Optical Materials, National Institute for Materials Science,\\[-0.8em]\small 1-1 Namiki, Tsukuba 305-0044, Japan\\
	\small$^{5}$Research Center for Materials Nanoarchitectonics, National Institute for Materials Science,\\[-0.8em]\small 1-1 Namiki, Tsukuba 305-0044, Japan\\
	\small$^{6}$Department of Physics, University of Konstanz, Konstanz 78464, Germany\\[1.8em]
	 \small$^\ast$Corresponding author. Email: bsoje@dtu.dk\\
	\small$^\ast$Corresponding author. Email: oded.zilberberg@uni-konstanz.de\\
	\small$^\dagger$These authors contributed equally to this work.
\end{center}




\subsection*{Materials and Methods}

\subsubsection*{Heterostructure assembly}
We fabricate the graphene heterostructures by exfoliating graphene and hBN onto SiO$_2$/Si substrates. Monolayer graphene is identified automatically via its optical contrast relative to the underlying substrate~\cite{jessen2018quantitative}. Assembly is done using a standard dry-transfer approach~\cite{wang2013one, pizzocchero_hot_2016,purdie_cleaning_2018} with a polycarbonate (PC) film supported on a hemispherical polydimethylsiloxane (PDMS) stamp. The stack is built by sequentially picking up and deliberately misaligning $\sim$30\,nm hBN, single-layer graphene, another $\sim$30\,nm hBN layer, and placing them onto a graphite flake roughly 5\,nm thick. The pickup is performed at around 110$^\circ$C, and the stack is released onto the target substrate by melting the PC layer at $\sim$190$^\circ$C.

\subsubsection*{Device fabrication}
Electron beam lithography was performed using an 80\,kV nanoBeam nB4 system. We employed a bilayer electron-beam resist consisting of 495K and 950K polymethyl methacrylate (PMMA) dissolved in anisole, resulting in approximate thicknesses of 200\,nm and 50\,nm, respectively. Each PMMA layer was baked at 180\,$^\circ$C for 30 minutes. Exposure was carried out at a beam current of approximately 2\,nA and an area dose of 1000\,$\mu$C/cm\textsuperscript{2}.\\
The samples were developed for 120\,s in a chilled (4\,$^\circ$C) solution of H\textsubscript{2}O:IPA (1:3), rinsed immediately with similarly cooled IPA, and dried under nitrogen flow. Prior to etching, a brief 8-second O\textsubscript{2} reactive ion etching (RIE) descum (10\,W, 80\,mTorr, 20\,sccm) was performed to remove PMMA residues, followed by a 90-second CF\textsubscript{4} etch (10\,W, 90\,mTorr, 70\,sccm) to selectively pattern the top hBN. The respective etch rates for PMMA were approximately 6.4\,nm/min for O\textsubscript{2} and 31.5\,nm/min for CF\textsubscript{4}, while the CF\textsubscript{4} plasma removed hBN at rates exceeding 400\,nm/min. Crucially, graphene layers serves as hard-masks for gente F-based etching \cite{jessen2019lithographic, son_atomically_2018, pizzocchero_hot_2016}, as confirmed by the absence of electrical breakdown in our devices, where bottom hBN serves as a reliable gate dielectric. A two-minute dummy run ("preconditioning") using the corresponding etch chemistry was performed before each plasma etch step to stabilize chamber conditions. For subsequent metallization steps, only the O\textsubscript{2} descum was applied.

\subsubsection*{Measurements}
Transport experiments are conducted in a Bluefors dilution refrigerator with a base temperature of 10\,mK at the mixing chamber. We perform two-terminal resistance measurements using Stanford Research SR860 lock-in amplifiers. A small AC voltage of 100\,$\mu$V is applied at a reference frequency of 17.777\,Hz to minimize pickup from ambient noise. To reduce high-frequency interference, commercial Q-Devil filters are mounted on the mixing chamber stage, contributing an additional 5\,k$\Omega$ series resistance, which has been subtracted from the data shown.

\subsubsection*{Numeric simulations}
\begin{figure} 
	\centering
	\includegraphics[width=1\linewidth]{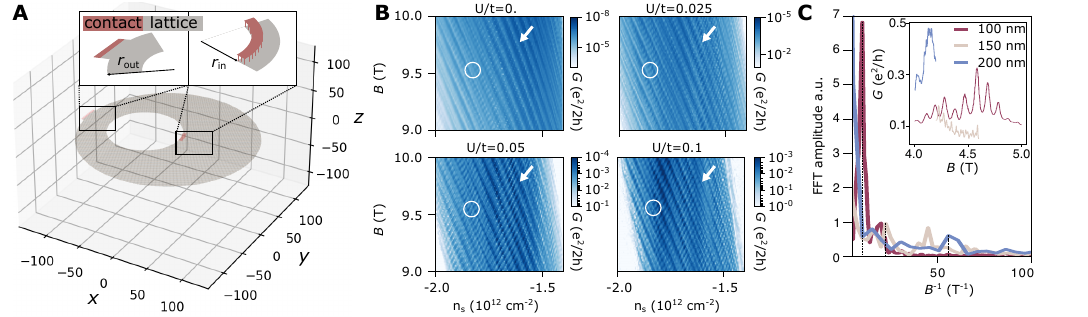}
	\caption{(\textbf{A}) We define a Corbino disk setup using the numerical package Kwant~\cite{groth_kwant_2014}. Two contacts are attached, one at the center from above (dot contact) and one laterally at the outer edge. (\textbf{B}). The two-terminal conductance is numerically computed as a function of magnetic field and carrier density for different disorder strengths. The arrow highlights the broadening of the spectral flow due to disorder, while the circled area marks regions where different spectral flows become indistinguishable. (\textbf{C}). Fourier transforms of the conductance are shown along fixed energies of $145$, $140$, and $144\,\mathrm{meV}$ for contacts of size $100$, $150$, and $200\,\mathrm{nm}$, respectively. Inset: two-terminal conductance plotted along the corresponding energy cuts.}
	\label{fig:S1}
\end{figure}

To numerically compute the two-terminal conductance through a Corbino disk system, we define a circular region on a honeycomb lattice with an outer radius of $r_\mathrm{out} = 120\,\mathrm{sites}$. The Corbino geometry is realized by removing a circular portion of the lattice with radius $r_\mathrm{in} = 50\,\mathrm{sites}$, representing the hole formed by the metallic contact potential. To probe conductance between the edges, we attach semi-infinite leads to the outer edge from the left and to the inner edge from the top, as illustrated in Extended Data Fig.~\ref{fig:S1}A. To enhance the conductance signal, the center of the hole is shifted by $35\,\mathrm{sites}$ toward the outer lead, reducing the distance between the two edges. This offset increases the current amplitude between the leads but does not alter the resonance condition given in Eq.\eqref{eq: edge states}. The resonances (with negative slope) correspond to the measured signal in the main text. They interfere with positive sloped signals arising from quantization of the edge states on the outer rim of the annulus. This is a finite-size effect arising from the numerics, which is not visible in the experiment.

To model a disordered system, we add a random onsite potential drawn from a normal distribution with zero mean and standard deviation $U/t$, where $t$ is the hopping amplitude in the graphene lattice. For $U/t = 0$, $0.025$, $0.05$, and $0.1$, the resulting conductance is shown in Extended Data Fig.~\ref{fig:S1}B. As indicated by the white arrows, the lighter blue stripes representing edge states broaden with increasing $U/t$, reflecting the smearing of spectral flows due to disorder. This effect arises because the measurement bandwidth $\Gamma$ increases with disorder strength, capturing the same flow over a wider energy range. Consequently, edge states from different valleys and spins appear as overlapping broadened features that cannot be resolved in experiments, as illustrated in the circled region of Fig.~\ref{fig:S1}B. In Fig.~\ref{fig:2}F, we present numerical data for $U/t = 0.025$.

\subsubsection*{Conductance along an energy cut}
To extract the period in the magnetic field for fixed energy, we use the relation
\begin{equation}\label{eq:Ecut}
    \nu=-\frac{\varepsilon_2^2}{v_F^2}\frac{2}{eB\hbar}-2-\frac{\phi_0}{B}\delta n_s\, ,
\end{equation}
where $\varepsilon_2$ is the energy of the electrons in the second bulk, and $\delta n_s=-0.2176 \times 10^{12}\,\mathrm{cm}^{-2}$ is the shift of the charge neutral point between the first and the second bulk~\cite{??}. By fixing the energy $\varepsilon_2$ of the second bulk, the relation in Eq.~\eqref{eq:Ecut} traces a curved line in the $\nu$–$B$ coordinate system, as shown in Fig.~\ref{fig:3}A. Extended Data Fig.~\ref{fig:S1}C presents the conductance along this selected energy for each device, along with the corresponding Fourier transforms. For devices with $100$ and $200\,\mathrm{nm}$ contacts, we observe frequency peaks at $10\,\mathrm{T}^{-1}$ and $56\,\mathrm{T}^{-1}$, which correspond to magnetic field periods of $\Delta B = 100\,\mathrm{mT}$ and $\Delta B = 18\,\mathrm{mT}$, respectively. For the contact with a radius of $150\,\mathrm{nm}$, two main peaks appear at $22.57\,\mathrm{T}^{-1}$ and $45.14\,\mathrm{T}^{-1}$. The first frequency matches the period predicted by Laughlin's argument, while the second, at twice the frequency, is attributed to a partial lifting of the spin-valley degeneracy in the inner edge state.
\subsubsection*{Extracting the group velocity of the edge states}
From the resonance condition depicted in Eq.~\eqref{eq: edge states} in the main text, we derive the expression of the spectral flow of the edge state labeled by the angular wavenumber $m$ in the energy-magnetic field coordinate as 
\begin{equation}\label{eq:spectral flow1}
    S_m^{\varepsilon}(B)=\Delta \varepsilon\left(m+\frac{B}{\Delta B}\right)\, .
\end{equation}
Using the relation between the filling and the energy $|\nu|=\frac{\varepsilon^2}{v_{F}^{2}}\frac{2}{eB\hbar}+2$, we obtain the expression of the spectral flow in the filling-magnetic field coordinate as
\begin{equation}\label{eq: stripe formula}
    S_{m}^\nu(B)=-\left[\frac{2\hbar}{R^2eB}\left(\frac{\tilde{v}_F}{v_F}\right)^2\left(m+\frac{\pi R^2}{\phi_0}B\right)^2-\delta n_s\frac{h}{eB}\right]-2\,.
\end{equation}
For each angular wavenumber $m$, the expression yields a spectral flow in the filling factor $\nu$ as the magnetic field $B$ varies. By fitting the spectral flow from Eq.~\eqref{eq: stripe formula} to the experimental data in Fig.~\ref{fig:3}A in the main text, we extract the angular wavenumber $m$, the group velocity $\tilde{v}_F$, and the radius $R$. The fitting procedure is conducted as follows: (i) we trace the coordinates along a stripe in the conductance from the experimental data, collecting them into a set of points $S_i = {(\nu_i^1, B_i^1), (\nu_i^2, B_i^2), \dots}$; (ii) for each stripe $S_i$, we select the points that lie in the gap, i.e., those satisfying $-15 < \nu_i < -13$; (iii) for all stripes within the range $4\,\mathrm{T} < B < 5\,\mathrm{T}$, we search for a series of consecutive integers $m$, along with a group velocity $\tilde{v}_F$ and radius $R$, such that the analytical pattern from Eq.~\eqref{eq: stripe formula} best matches the extracted data points from all $S_i$; for the data in Fig.~\ref{fig:3}A in the main text, we obtain $m = {-11, -10, \dots, -3}$, $v_F / \tilde{v}_F = 1.86$, and $R = 118\,\mathrm{nm}$, see Fig.~\ref{fig:3}D in the main text; (iv) for spectral flows that lie deep within the gap and near the Landau levels, we fix the angular wavenumber $m$ and fit only $\tilde{v}_F$ and $R$. The growing deviation between the fitted lines and experimental data is attributed to the non-linear dispersion of edge states as they approach the Landau levels.

\clearpage 





\end{document}